\documentclass[12pt]{article}
\usepackage{graphicx}
\usepackage{cite}

\newcommand\pubnumber{DPF2013-73}
\newcommand\pubdate{\today}

\def\fsu{Florida State University\\
Department of High Energy Physics, Tallahassee, FL 32306}
\def\support{\footnote{This research funded under DOE Grant: \uppercase{DE}-\uppercase{SC}0010102}}

\def\Title#1{\begin{center} {\Large #1 } \end{center}}
\def\Author#1{\begin{center}{ \sc #1} \end{center}}
\def\Address#1{\begin{center}{ \it #1} \end{center}}

\newcommand\pubblock{\rightline{\begin{tabular}{l} \pubnumber\\
         \pubdate  \end{tabular}}}
\newenvironment{Abstract}{\begin{quotation}  }{\end{quotation}}
\newenvironment{Presented}{\begin{quotation} \begin{center} 
             PRESENTED AT\end{center}\bigskip 
      \begin{center}\begin{large}}{\end{large}\end{center} \end{quotation}}

\def\beq{\begin{equation}}
\def\eeq#1{\label{#1}\end{equation}}
\def\eeqn{\end{equation}}

\def\beqa{\begin{eqnarray}}
\def\eeqa#1{\label{#1}\end{eqnarray}}
\def\eeqan{\end{eqnarray}}

\let\bar=\overbar

\def\Dslash{\not{\hbox{\kern-4pt $D$}}}
\def\dslash{\not{\hbox{\kern-2pt $\del$}}}

\def\alphas{\alpha_s}
\def\msb{{\bar{\ssstyle M \kern -1pt S}}}

\begin{document}
\begin{titlepage}
\pubblock

\vfill
\Title{Threshold Resummation and Determinations of Parton Distribution Functions}
\vfill
\Author{ David Westmark\support}
\Address{\fsu}
\vfill
\begin{Abstract}
 The precise knowledge of parton distribution functions (PDFs) is indispensable
 to the accurate calculation of hadron-initiated QCD hard scattering observables.
 Much of our information on PDFs is extracted by comparing deep inelastic scattering
 (DIS) and lepton pair production (LPP) data to convolutions of the PDFs with the
 partonic cross sections of these processes.
 It is known that partonic cross sections receive large corrections in regions of phase
 space near partonic thresholds that can be resummed using threshold resummation techniques.
 The effect of threshold resummation on DIS and LPP differs because partonic thresholds
 for the two processes occur in different kinematic regions.
 Recent global fits for PDFs have included DIS data from the large Bjorken $x$ and
 moderate $Q^2$ region where threshold effects have been shown to be large.
 The present project explores the effects of simultaneously incorporating threshold
 resummation in both DIS and LPP and to evaluate the effects of such additions on global
 fits.
\end{Abstract}
\vfill
\begin{Presented}
DPF 2013\\
The Meeting of the American Physical Society\\
Division of Particles and Fields\\
Santa Cruz, California, August 13--17, 2013\\
\end{Presented}
\vfill
\end{titlepage}
\def\thefootnote{\fnsymbol{footnote}}
\setcounter{footnote}{0}

\section{Introduction: Inclusive Cross Section}

A differential inclusive cross section for a process $H1+H2 \rightarrow A+X$ with observed
final state $A$ takes the generic form of a convolution of hard scattering cross sections and parton distribution
functions,
\beq
\mathrm{d}\sigma(x,Q^2)=\sum_{a,b}\mathrm{d}x_{1}\,\mathrm{d}x_{2}\,\mathrm{d}z\,f_{a/H_{1}}(x_{1},Q)f_{b/H_{2}}(x_{2},Q)\mathrm{d}\hat{\sigma}_{ab}(z,Q^2)\delta(x-x_{1}x_{2}z),
\eeqn
where $\hat{\sigma}_{ab}$ is the hard scattering cross section describing the interactions of two
partons $a$ and $b$, and $f_{a/H}$ is a parton distribution function (PDF).
The PDFs can be interpreted as the probability of obtaining a parton $a$ from a hadron $H$ at an energy scale $Q$
and momentum fraction of the parent hadron between $x$ and $x+dx$.
Unlike $\hat{\sigma}$, which is calculated via a perturbative expansion in $\alphas$,
PDFs cannot be calculated perturbatively and are determined by comparing data from
hadron-initiated processes to the theoretical predictions of the data.
PDFs have thus far been successfully inferred from data with fixed-order
techniques\cite{Alekhin:2012ig,Ball:2011uy,Lai:2010vv,JimenezDelgado:2008hf,Martin:2009iq,Owens:2012bv}
using global sets of data that include deep inelastic scattering (DIS) and lepton pair production (LPP).
The goal of this study is to qualitatively demonstrate that the inclusion of threshold
resummation, an all-orders technique, can impact nucleon PDFs.

\section{Threshold Resummation}

Logarithmic corrections appear in the perturbative expansion of $\hat{\sigma}$ that can become
large in threshold regions, when the leading order process takes all available energy
and final state gluons are soft, potentially spoiling the perturbative series.
In this situation, an all-orders resummation of these corrections is needed to
preserve the perturbative series\cite{Catani:1989ne,Sterman:1986aj}.
These threshold logarithms appear at all orders of perturbation theory beyond LO in a
predictable manner, but are convoluted with the PDFs and cannot be easily resummed.
Threshold resummation is therefore naturally performed in a Mellin
conjugate space with Mellin moments $N$, where the convolution integrals are reduced to simple products.
In this space, the threshold regions manifest as $N\rightarrow\infty$ and the threshold
logarithms as powers of $\ln(N)$.
It has been shown that when the threshold logarithms are resummed, one obtains an
exponent\cite{Catani:1989ne,Sterman:1986aj,Kidonakis:1997gm}:
\beq
C^{\it res}_N (\alphas) = g_0(\alphas) \exp(G_N(\alphas)),
\eeqn
where $g_0$ is a series in $\alphas$ that is finite as $N\rightarrow\infty$ and $G_N$ contains the
resummed threshold logarithms.
Resummation corrections are known exactly for many processes up to next-to-next-to-leading
logarithmic order (NNLL)\cite{Vogt:2000ci,Catani:2003zt}, and $\mathrm{N}^{3}$LL\cite{Moch:2005ba}
corrections are partially known for some processes.

One must invert the resummation corrections from Mellin space in order to compare them
with data, a task that requires a method for avoiding the Landau pole divergence.
One of the more common methods, and the one used in this study, is the minimal prescription\cite{Catani:1996yz}, which
simply chooses an inversion integral path to the left of the Landau pole, but to the right of all
other poles.
By subtracting the expansion of the resummation exponent to an appropriate
order in $\alphas$, one can also match the resummed corrections to fixed-order calculations.
For the purposes of this study, it is sufficient to consider an NLO fixed-order calculation with
matched NLL corrections, meaning that logarithmic terms of the order $\alphas^m\ln^{m+1}(N)$ and
$\alphas^m\ln^{m}(N)$ are fully resummed.

\section{Deep Inelastic Scattering and Lepton Pair Production}

DIS and LPP data are often used in global fits of PDFs to constrain the lighter quark and
antiquark distributions.
DIS data is a primary source of information on the valence quarks, particularly the up quark,
since $F_2$, the familiar DIS structure function, behaves as $F_2\sim(4u+d)$ at high $x$.
The comparison of proton-proton and proton-neutron LPP data provides constraints to the
$\bar{u}-\bar{d}$ distribution and the $\bar{d}/\bar{u}$ ratio.

\begin{figure}[htb]
\centering
\includegraphics[height=2in]{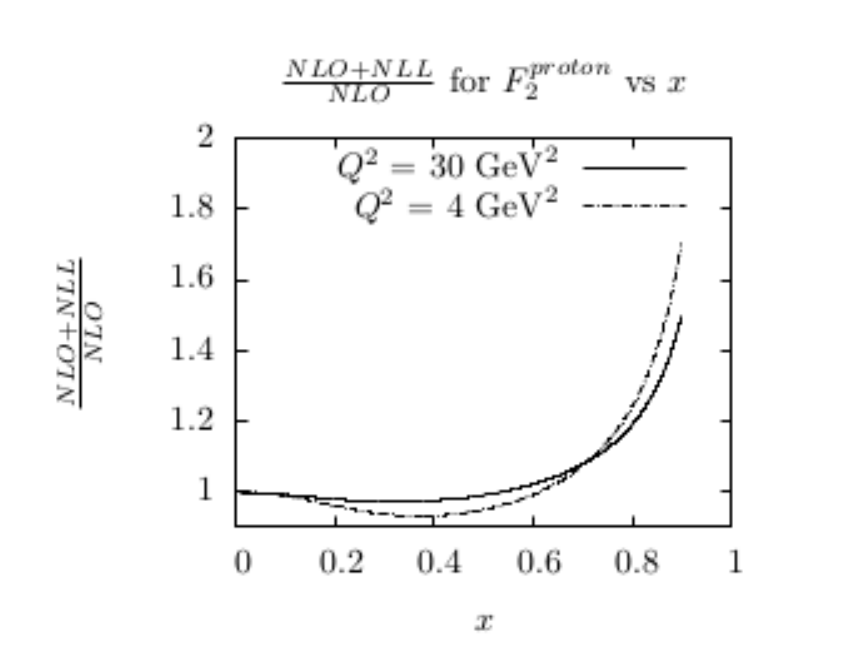}
\includegraphics[height=2in]{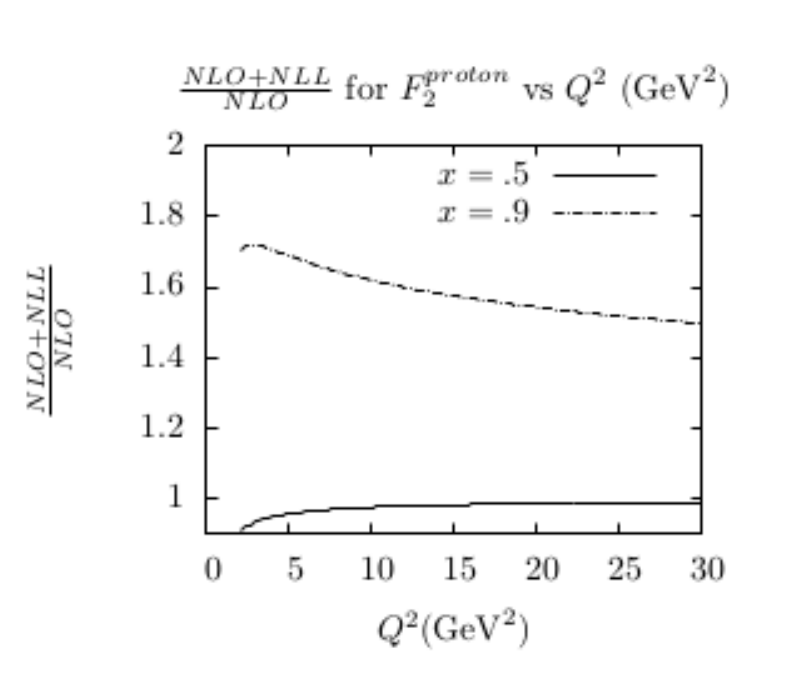}
\caption{Left Figure: Ratio of NLL+NLO to NLO vs $x$ for $F_2^{proton}$ with $Q^2$ = 4 $\mathrm{GeV}^2$
and $Q^2$ = 30 $\mathrm{GeV}^2$ (using CTEQ6M PDFs\cite{Pumplin:2002vw} and all scales set equal to $Q$).
Right Figure: Ratio of NLL+NLO to NLO vs $Q^2$ $\mathrm{GeV}^2$ for $F_2^{proton}$ with $x=.5$
and $x=.9$ (using the same PDFs and choices of scale).}
\label{fig:disnorm}
\end{figure}

DIS is a single hadron process initiated by vector boson exchange between a lepton and a parton
inside the hadron (schematically $l+H\rightarrow l'+X$).
The hadronic final state $X$ has an invariant mass squared of $W^2=M^2+Q^2(1/x-1)$, where
$M^2$ is the squared nucleon mass, $Q^2$ is the negative of the invariant
mass squared of the exchanged boson in the process, and $x$ is the Bjorken scaling variable.
Threshold for DIS corresponds to $W^2=M^2$, when all final state gluons are soft, and
therefore occurs at $x=1$.
In the left side of Figure~\ref{fig:disnorm}, it is demonstrated that the threshold resummation corrections are
indeed greater at high x.
The right side of Figure~\ref{fig:disnorm} shows the $Q^2$ dependence of the resummation corrections.
It can be seen that this high x behavior is consistent across this range of $Q^2$.
Therefore, the inclusion of data sets in these kinematic regions\cite{Whitlow:1991uw,Malace:2009kw} will
impact global fits of PDFs.

LPP is a hadron-hadron scattering event with center of mass energy squared $S$, where a parton from one hadron
interacts with a parton from the other hadron to produce a massive vector boson that decays into a lepton pair
of invariant mass squared $Q^2$.
In LPP, the threshold region corresponds to $Q^2$ taking all the available energy of
the partonic system, $\hat{S}=x_1x_2S$ with $x_1$ and $x_2$ being the parton's momentum fraction,
so that any final state gluons are soft.
Therefore, the threshold region occurs when $\hat{\tau}=Q^2/\hat{S}=1$.
In some LPP observables, one integrates across all available momentum fractions
$x_1$ and $x_2$, implying that threshold conditions can be reached for any value of $Q^2/S$.

\begin{figure}[htb]
\centering
\includegraphics[height=2in]{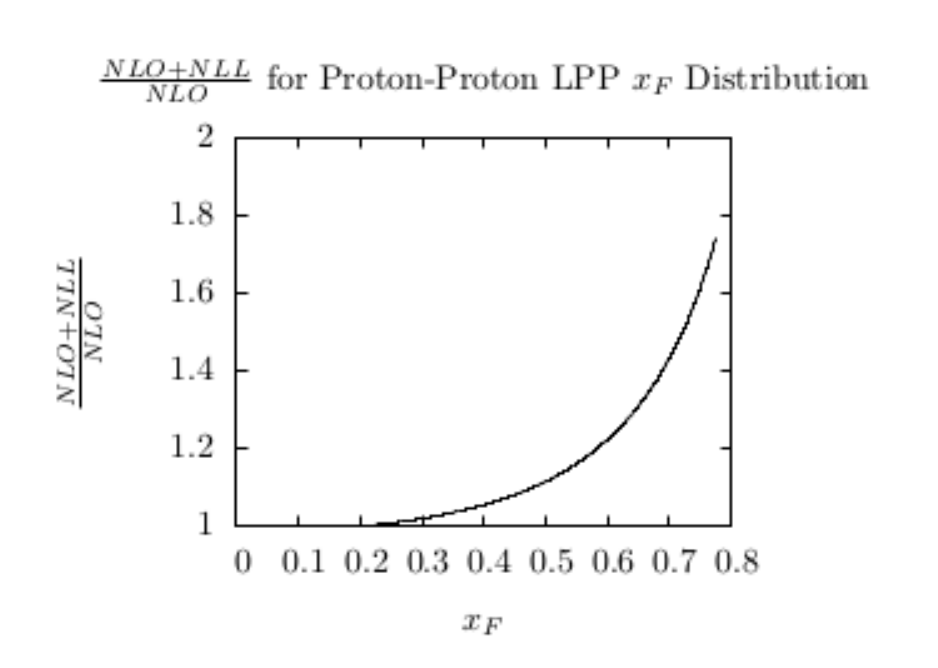}
\caption{$\frac{\mathrm{d}^2\sigma}{\mathrm{d}Q\mathrm{d}x_F}$ at NLL+NLO normalized to NLO with
$Q$ = 8 $\mathrm{GeV}$, $\sqrt{S}$ = 38.76 $\mathrm{GeV}$ (using the same PDFs and scale choices as in Figure~\ref{fig:disnorm}).}
\label{fig:lppnorm}
\end{figure}

LPP data is often a doubly-differential cross section in both $Q^2$ and $x_F=\frac{2p_L}{\sqrt{S}}$
(or $Y$, the lepton pair's rapidity), where $p_L$ is the longitudinal momentum of the lepton pair.
In Figure~\ref{fig:lppnorm}, the resummation corrections to $\frac{\mathrm{d}^2\sigma}{\mathrm{d}Q\mathrm{d}x_F}$ are shown.
Here, $Q$,$\sqrt{S}$, and the range of $x_F$ were chosen to match E866/NuSea\cite{Hawker:1998ty,Webb:2003ps}
conditions to demonstrate that resummation corrections are sizable in kinematic regions with LPP data.
Notice that the resummation effects grow with increasing $x_F$;
this is because at higher values of $x_F$, one of either $x_1$ or $x_2$ becomes large,
and the steeply falling PDFs at higher $x$ push the fixed-order calculation toward threshold.
Therefore, the resummation corrections become the most dominant contribution to the observable.

\section{Conclusions}

Recent global fits of PDFs\cite{Alekhin:2012ig,Owens:2012bv} have included data from threshold regions, where
threshold resummation corrections to DIS and LPP have been shown to be sizable.
In addition, these corrections occur in different kinematic regions between DIS and
LPP, and their simultaneous inclusion in global fits will affect different PDFs.
The valence PDFs at large $x$ and moderate $Q^2$ will be particularly affected by the
DIS resummation corrections, and we should expect to see more constraints in this region.
This implies that it may be necessary to include threshold resummation in global fits
of PDFs in order to gain a more accurate understanding of the PDFs. To this end, a PDF
fitting program has been modified to include threshold resummation corrections at
NLO+NLL for DIS and LPP, and preliminary global fits have been performed. The results of
this study are expected to be published at a later date.

\end{document}